\documentclass[useAMS,usenatbib]{mn2e}

\usepackage{graphicx}
\usepackage{txfonts}
\usepackage{natbib}

\title[An unexpected outburst from A0535+262]{An unexpected outburst from A0535+262}

\author[A. B. Hill]{A. B. Hill$^{1}$\thanks{E-mail:abh@astro.soton.ac.uk} 
, A. J. Bird$^{1}$, A. J. Dean$^{1}$, V. A. McBride$^{1}$, V. Sguera$^{1}$\thanks{New Address: INAF/IASF Bologna, via Piero Gobetti 101, 40129 Bologna, Italy}, 
\newauthor
D. J. Clark$^{1}$, M. Molina$^{1}$, S. Scaringi$^{1}$ and S.E. Shaw$^{1}$\\
$^{1}$School of Physics \& Astronomy, University of Southampton, UK\\}
\begin{document}

\date{Accepted/Received}

\pagerange{\pageref{firstpage}--\pageref{lastpage}} \pubyear{2007}

\maketitle
\label{firstpage}

\begin{abstract}
A0535+262 is a transient Be/X-ray binary system which was in a
quiescent phase from 1994--2005.  In this paper we report on the
timing and spectral properties of the INTEGRAL detection of the source
in October 2003.  The source is detected for $\sim$6000 seconds in the
18--100 keV energy band at a luminosity of
$\sim$3.8~$\times$~$10^{35}$ergs~s$^{-1}$; this is compatible with the
high end of the range of luminosities expected for quiescent emission.
The system is observed to be outside of the centrifugal inhibition
regime and pulsations are detected with periodicity,
P$=$103.7~$\pm$~0.1 seconds.  An examination of the pulse history of
the source shows that it had been in a constant state of spin-down
since it entered the quiescent phase in 1994.  The rate of spin-down
implies the consistent presence of an accretion disk supplying torques
to the pulsar.  The observations show that the system is still active
and highly variable even in the absence of recent Type I or Type II
X-ray outbursts.
\end{abstract}


\section{Introduction}
A0535+262 is a high mass X-ray binary (HMXB) system comprising a
neutron star primary and an O9.7IIIe donor star
\citep{Giangrande1980}.  A0535+262 is an example of the Be X-ray binary
subclass.  Be/X-ray binaries show two types of X-ray outbursts which
are regularly referred to as Type I and Type II \citep{Stella86}.
Type I outbursts are associated with the time of periastron passage as
the neutron star passes closest to the Be donor.  The X-ray luminosity
is typically $10^{36}$--$10^{37}$ergs~s$^{-1}$ and they last for
several days.  Type II outbursts can occur at any phase of the orbit,
they have X-ray luminosities of $>10^{37}$ergs~s$^{-1}$ and they can
last for several weeks \citep{Negueruela1998b}.

A0535+262 was discovered by Ariel V during a large Type II outburst in
1975 \citep{Coe1975, Rosenberg1975}.  Since then the source has been
observed to undergo numerous outbursts, however there were no reported
detections of X-ray outburst activity from 1994--2005 \citep{Coe2006,
Kretschmar2005}.  The source reappeared in a Type II outburst in
May/June 2005 and was detected by Swift \citep{Tueller2005} and RHESSI
\citep{Smith2005}.  It was subsequently seen to undergo a Type I
outburst in August 2005 \citep{Kretschmar2005, Caballero2007}.

The orbital parameters of the system were first determined by
\citet{Finger96} following the 1994 outburst:
\begin{itemize}
\item P$_{\rm{orbital}}$~=~110.3~$\pm$~0.3 days 
\item a$_x$$\sin{i}$~=~267~$\pm$~13 lt-sec
\item \emph{e}~=~0.47~$\pm$~0.02  
\end{itemize}
The orbital ephemeris, which corresponds to the time of periastron,
and the orbital period have been more recently estimated by
\citet{Coe2006}: 
\begin{itemize}
\item $\tau_{\rm{periastron}}$~=~2450094~$\pm$1 JD
\item P$_{\rm{orbital}}$~=~110.0~$\pm$~0.5 days  
\end{itemize}
Measurements of the cyclotron features by RXTE and INTEGRAL during the
2005 outburst estimated the magnetic field strength of the neutron
star as $\sim$4~$\times$~10$^{12}$~Gauss \citep{Caballero2007}.  A
distance to the source of 2~kpc was estimated by \citet{Steele98}
through the observation of the spectral type and the reddening of the
optical counterpart of A0535+262.

A number of observations have been made of A0535+262 whilst it is in
the quiescent phase.  EXOSAT observations were made in 1985--1986
between outbursts by \citet{Motch91}; pulsations were detected and a
luminosity of 0.7--1.4~$\times$~$10^{35}$~ergs~s$^{-1}$ for a source
at 2 kpc was measured.  During the 1994--2005 quiescent phase the
system was observed by the RXTE-PCA instrument in 1998 and by BeppoSAX
in 2000 \citep{Negueruela2000, Mukherjee2005}; both telescopes
reported luminosities of
$\sim$1.5--4.5~$\times$~$10^{33}$~ergs~s$^{-1}$ which indicated the
system was in the centrifugally inhibited regime despite both
telescopes detecting pulsations in some of their observations.  In
this paper we present soft $\gamma$-ray observations of A0535+262
during a flare whilst the source was in a quiescent state.  A timing
and spectral analysis is performed on the INTEGRAL data and the
results interpreted in the context of previous observations.  The long
term pulse history of the source is discussed.

\section{INTEGRAL data}
Since its launch in October 2002, INTEGRAL (the INTErnational
Gamma-Ray Astrophysics Laboratory) has been performing regular scans
of the Galactic plane.  The IBIS imaging instrument is sensitive in
the 15 keV -- 10 MeV energy range \citep{ibis} and includes the
INTEGRAL Soft Gamma-Ray Imager (ISGRI).  Recently, the IBIS Survey
Team released the 3$^{rd}$ IBIS/ISGRI soft $\gamma$-ray survey
catalogue comprising 40 Ms of data from the first 3.5 years of core
and public program observations \citep{cat3}.  Included in this
catalogue was the first announced detection of A0535+262 by the IBIS
Survey Team; they observed that it was seen at 9.4~$\sigma$ in their
all-sky mosaic with an average flux of 3.0~$\pm$~0.3 mCrab in the
20--40 keV energy band and an exposure of $\sim$400 ks.  However, the
data set used by the IBIS survey team did not include any observations
of A0535+262 during the 2005 outbursts.

As A0535+262 is known to be a transient source we searched back
through the INTEGRAL data archive to identify any outbursts in the
catalogue 3 data set.  The INTEGRAL data is organised in short
pointings, known as science windows, of approximately 2000\,s.  The
source was only significantly detected in 3 science windows during
INTEGRAL orbit number 127; this corresponds to the source being
detected on 28 October 2003 from 09:45:10 to 12:39:54 UTC.  Prior and
post to this the source was outside of the field of view of the
telescope.  During the observation and afterwards there was increased
solar activity; an X17 solar flare on October 28 and an X10 flare on
October 29 \citep{rhessi_flare}.  The flare on October 28 2003 began
at 09:41 UT, was at maximum at 11:10 UT and ended at around 11:24 UT;
SPI, the spectrometer onboard INTEGRAL, showed a strong increase in
counting rate at $\sim$11:02 \citep{spi_flare}.  Consequently there is
significant background noise, especially in the latter half of the
observation.  After the end of the observation there are no further
observations of this field.

\section{Analysis}
\begin{figure}
\centering
\includegraphics[width=0.95\linewidth, clip]{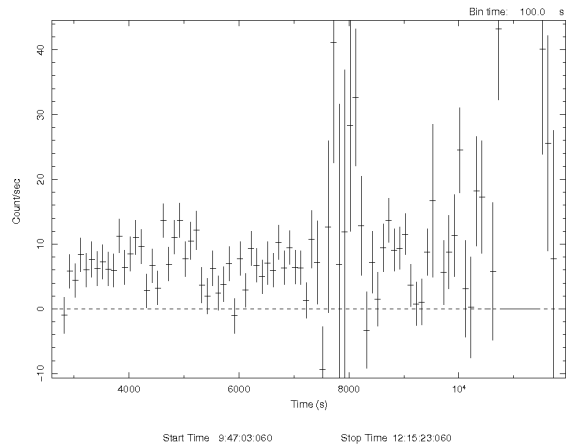}
\caption{The IBIS/ISGRI 20-40 keV 100~s binned light curve of A0535+262 on 28 October 2003.  The poor data quality in the latter half of the light curve is attributed to contamination from solar flare activity.}
\label{fig:lc}
\end{figure}

The IBIS/ISGRI data was analysed using the \emph{Offline Standard
Analysis, OSA,} software version 5.1.  The 20--40 keV IBIS/ISGRI light curve
of A0535+262 is shown in Figure~\ref{fig:lc}.  It is clear that the
quality of the data during the latter half of the observation has been
compromised by the increased background resulting from the solar flare
activity.  This is also evident in the increasing amount of instrument
dead time throughout the observation.  Figure~\ref{fig:lc} does not
show any general trend with time such that there is no indication
whether the source is beginning or ending an outburst.  The lack of
data prior to and post of the observation means that we cannot place
any real limits on the outburst duration.  The IBIS/ISGRI mosaic of the
three science windows detected A0535+262 with a flux of 62~$\pm$~3
mCrab (a 19~$\sigma$ detection) in the 20--40 keV band and a
32~$\pm$~5 mCrab (a 6~$\sigma$ detection) in the 40--60 keV energy
band.

An investigation of the RXTE-ASM light curve of A0535+262 shows no
observations of the source were made between 04:48--19:12 UTC on 28
October 2003.  The few observations made prior and post of these times
show no clear indication of any outbursting behaviour.  The sparsity
of the ASM data around the INTEGRAL observation is potentially due to
the high solar activity.

\subsection{Timing Analysis}
\begin{figure}
\centering
\includegraphics[width=0.95\linewidth, clip]{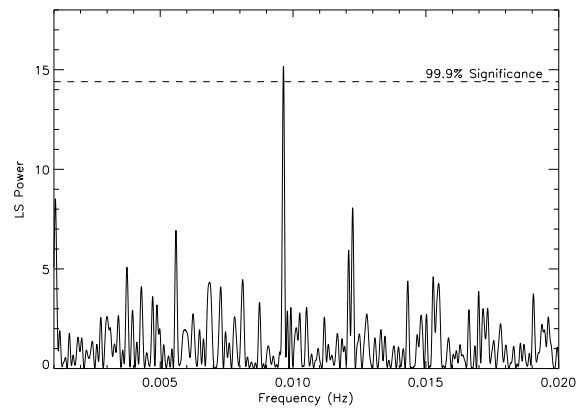}
\caption{The Lomb-Scargle periodogram of the 1\,s binned light curve of A0535+262.  The dashed line represents the 99.9$\%$ significance level.  There is a clear spike at $\sim$9.64 ~$\times$~ 10$^{-3}$ Hz. }
\label{fig:period}
\end{figure}

A light curve with 1 second binning was generated using the \emph{OSA
ii\_light} tool in the 20--40 and 40--60 keV energy bands.  The light
curves were barycentre corrected for the orbital motion of the
INTEGRAL satellite using the \emph{OSA barycent} command.  The 1s
binned 20--40 keV light curve was analysed for any periodic signals
using the Lomb-Scargle periodogram method by means of the fast
implementation \citep{Scargle82, Press89}.  The resulting power
spectrum is shown in Figure~\ref{fig:period}; a peak is clearly
evident at 0.00964 Hz with a power of $\sim$15.1.  This corresponds to
a period of 103.7 seconds; A0535+262 has been observed to have a spin
period of between 103--104 seconds \citep{Mukherjee2005}.

Having confirmed the INTEGRAL source as A0535+262 the light curve was
corrected for the orbital motion of the neutron star around the donor
using the orbital parameters of \citet{Finger96} and the outburst
ephemeris of \citet{Coe2006}.  The observation occured at an orbital
phase 0.81~$\pm$~0.07 when the neutron star was moving perpendicular to the line
of sight.  The timing analysis was then re-performed and a series of
Monte-Carlo methods used to verify the periodicity detection.  The
spin period is measured to be 103.7~$\pm$~0.1 seconds; the period
error is calculated using the method of \citet{Horne86} and is
confirmed using the bootstrap method of \citet{Kawano95}.  To confirm
the significance of the Lomb-Scargle peak a randomization test was
performed \citep{Hill05}.  The 1 second binned light curve was
randomised and the resulting periodogram inspected; 200,000 light
curves were simulated and indicated that a Lomb-Scargle power of 15.1
represented a 5.2$\sigma$ detection of these pulsations.

\subsection{Spectral Analysis}

\begin{figure}
\centering
\includegraphics[width=0.95\linewidth, clip]{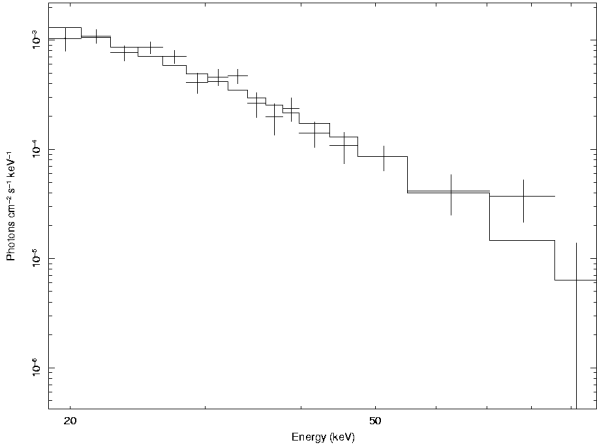}
\caption{The unfolded energy spectrum of the ISGRI A0535+262 data.  The spectrum has been fitted with a thermal bremsstrahlung model (see Table~\ref{tab:spec}).}
\label{fig:spec}
\end{figure}

Using the standard \emph{OSA} software, a spectrum was extracted from
the IBIS/ISGRI data.  Unfortunately, A0535+262 was outside of the field of
view of the INTEGRAL X-ray monitor, JEM-X, and hence only the IBIS/ISGRI
spectrum was available for analysis.  The 18-100 kev IBIS/ISGRI spectum was
fit using the \emph{XSPEC} software, version 12.3, a systematic error
of 2\% was included. The best fit was obtained by an absorbed thermal
bremmstrahlung model ($\chi^{2}_{\nu}$~=~1.17) with kT=23$^{+6}_{-4}$
keV (see Figure~\ref{fig:spec}), however an absorbed simple power law
provided a reasonable fit ($\chi^{2}_{\nu}$~=~1.24). In both cases the
absorption was fixed to the expected Galactic line-of-sight column
density of 0.6~$\times$~$10^{22}$cm$^{-2}$.  The results of the fits
are shown in Table~\ref{tab:spec}.  The 18--100 keV flux for the
source is $\sim$7.9~$\times$~$10^{-10}$ergs~cm$^{-2}$~s$^{-1}$.
 
\begin{table}
\centering
\caption{The results of the spectral fits to the ISGRI A0535+262 data in the 18--100 keV band.}
\begin{tabular}{|c|cc}
\hline
Parameter & Power law & Thermal Bremsstrahlung \\
\hline
Normalisation	& 8.0$^{+13}_{-5}$ & 0.25$^{+0.08}_{-0.06}$\\
Photon index	&  2.9$\pm$0.3 & - \\
kT		& -	&	23$^{+6}_{-4}$ keV\\
$\chi$$^{2}$~/~dof	&	31.1/25	& 29.2/25\\
\end{tabular}
\label{tab:spec}
\end{table} 
 
\section{Discussion}
It had been believed that the A0535+262 system had been in a period of
inactivity for the 10 years prior to the 2005 outburst
\citep{Coe2006}.  In the 1994--2005 period a number of X-ray
observations were made of the system whilst it was in quiescence.
These include the observations made by the RXTE-PCA instrument in
August and November 1998 which detected the source at a luminosity of
2.0--4.5~$\times$~10$^{33}$ergs s$^{-1}$ \citep{Negueruela2000}.  The
luminosity of the source at this time was so low as to indicate that
the source had entered the centrifugal inhibition phase despite the
detection of pulsations.  Observations by BeppoSAX in September and
October 2000 and March 2001 also observed the source during a state of
very low luminosity, 1.5--4.0~$\times$~10$^{33}$ergs s$^{-1}$ and
again detected pulsations \citep{Mukherjee2005}.

Despite a significant level of exposure on this region of the sky,
$\sim$400 ks, IBIS/ISGRI only detects A0535+262 on one occasion for a
duration of $\sim$6 ks at a level of $\sim$62 mCrab in the 20--40
keV energy band; this indicates that IBIS/ISGRI observed some level of
flaring behaviour.  No other X-ray instrument reports a detection at
this time, however, this may be attributable to the high level of
solar activity which occurred.  The time of the IBIS/ISGRI detection
corresponds to a phase of 0.81~$\pm$~0.07 using the ephermeris of
\citet{Coe2006} and orbital period of \citet{Finger96}; this is $\sim$21 days prior to periastron and is somewhat early to be a Type I outburst.  However, the calculated phase is within 3$\sigma$ of periastron and Be X-ray binary systems have been known to undergo Type I outbursts prior to and post periastron.  EXO 2030+375, for example, has been seen to outburst between 4 days prior to and 6 days post periastron (phase 0.91 and 0.09 respectively) in its $\sim$46 day orbit \citep{Wilson2002}.  This shift is explained by a density perturbation in the Be star's equatorial disk.  \citet{Coe2006} show that the H$\alpha$ profiles of A0535+262 in Dec 2002 and Mar 2005 exhibit some asymmetry in the double peaked structure which may indicate a density perturbation in the Be disk; density perturbations in the disk of A0535+262 have previously been reported by \citet{Negueruela1998a}.  However, in 2003 at the time of the INTEGRAL observation, \citep{Coe2006} estimate the Be disk size to be at the 7:1 resonance radius of the model of \citet{Okazaki2001}.  For a Type I outburst to occur the neutron star must accrete matter through Roche lobe overflow, this is possible when the circumstellar disk reaches the 4:1 resonance radius, which in the case of A0535+262 did not happen until early 2005 \citep{Okazaki2001, Coe2006}.

It is possible that the activity observed by IBIS occurs each periastron passage and that it has not previously been noticed because it is short lived and requires a high sensitivity instrument to be be looking at it.  Looking through the INTEGRAL archive we note that A0535+262 has been observed on numerous other occassions by INTEGRAL between 2003 and 2005 and in two additional instances the pointings coincided with a periastron passage, however in neither of these instances was a significant detection of A0535+262 made by INTEGRAL.  Analysis of the RXTE All Sky Monitor (ASM) X-ray data during 10 years of quiescence of A0535+262 by \citet{Coe2006} find a modulation in the X-ray flux on the orbital period with maximum flux at periastron. 

\begin{figure*}
\centering
\includegraphics[width=0.8\linewidth, clip]{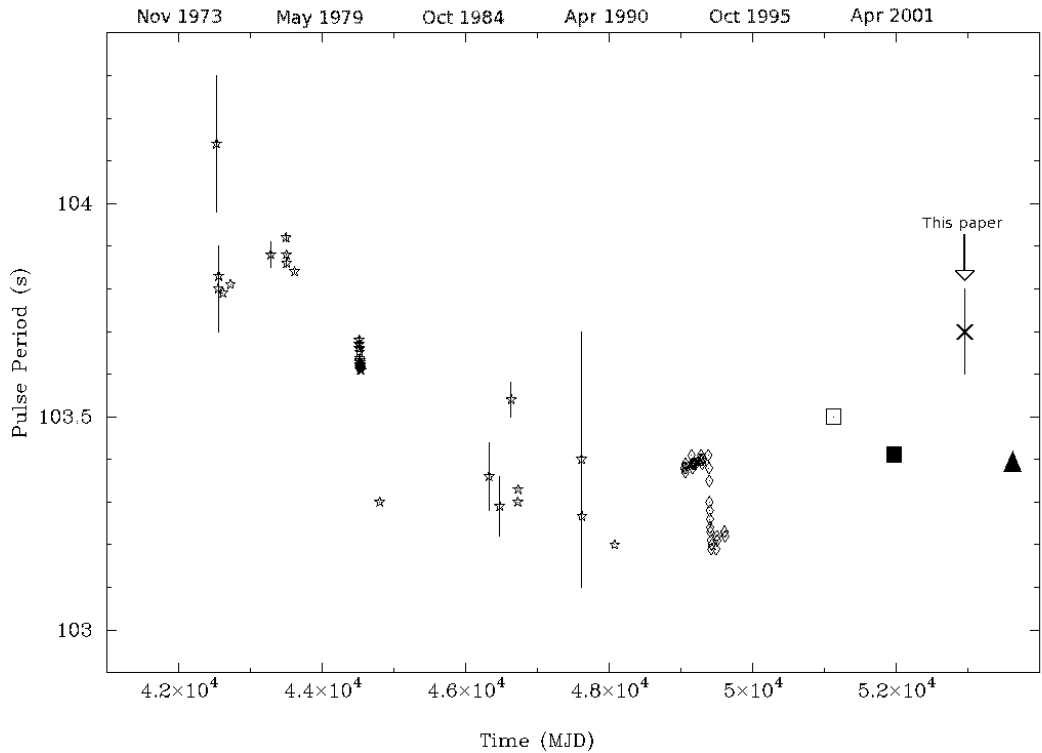}
\caption{The long term pulse period history of A0535+262 adapted from \citet{Mukherjee2005}: the diamonds represent the BATSE measurements of \citet{Finger96}; the open square indicates the RXTE-PCA measurement by \citet{Negueruela2000}; the filled square indicates the BeppoSAX measurement of \citet{Mukherjee2005}. The cross indicates the INTEGRAL-IBIS/ISGRI measurement at MJD 52940.4 (this paper). The triangle represents the INTEGRAL-IBIS/ISGRI measurement of \citet{Caballero2007} at MJD 53615.5137.}
\label{fig:pp_hist}
\end{figure*}

The flux of the source is measured as
7.9~$\times$~$10^{-10}$ergs~cm$^{-2}$~s$^{-1}$ in the 18--100 keV
band.  Taking a distance of 2 kpc \citep{Steele98} implies a
luminosity of $\sim$3.8~$\times$~$10^{35}$ergs~s$^{-1}$.  This
luminosity is at the lower end of what could be expected of a Type I
outburst which have typical luminosities of
$\sim$10$^{36}$ergs~s$^{-1}$ and is vastly under-luminous for a Type
II outburst which is typically in excess of 10$^{37}$ergs~s$^{-1}$.
Earlier estimates of the source distance range from 1.3 to 2.4 kpc
\citep{Steele98}; within this range the source is still underluminous
for either Type I or Type II bursts.  Additionally, Type II outbursts
typically last for approximately a month, consequently we would expect
detections by other instruments despite the high solar activity if
this were an example of a Type II outburst.  Whilst the luminosity of
the source is compatible with a Type I outburst the time of outburst
does not correspond to the periastron passage.  As discussed above this may suggest that the source activity is not attributable to a
Type I outburst.  We suggest that the INTEGRAL detection may have been a brief
flare during the quiescent phase.  This could potentially have been
triggered by an increased density of the material thrown out by the
stellar wind of the Be star.

Assuming an efficiency $\eta$=1 in the conversion of gravitational energy into luminosity then we can calculate the mass accretion rate: 
\begin{eqnarray*}
\dot{m} &=& 2.0\times10^{15}~{\rm g~s^{-1}} \\
&=& 3.2\times10^{-11}~{\rm M_{\sun}~yr^{-1}}
\end{eqnarray*}
The magnetic field strength of the neutron star in A0535+262 is
$\sim$4~$\times$~10$^{12}$ Gauss \citep{Caballero2007}.  For the known
magnetic field strength, the mass accretion rate and the spin period
of the pulsar the corotation radius is estimated using the formulation
of \citet{Stella86}
\begin{eqnarray*}
r_{co} & \sim & 3.8\times10^9~{\rm cm}
\end{eqnarray*}
and the magnetospheric radius is estimated using the formulation of \citet{Frank2002}
\begin{eqnarray*}
r_{m} &\sim& 1.5\times10^9~{\rm cm}
\end{eqnarray*}
The corotation radius defines the distance at which the neutron star
rotation velocity is equal to the Keplerian velocity.  The
magnetospheric radius defines the region in which the magnetic field
strongly effects the dynamical properties of infalling material.  If
the magnetospheric radius is larger than the corotation radius
material is then accreted material is stopped at the magntospheric
boundary and may be ejected beyond the accretion radius via the
propeller effect; this is known as the centrifugal inhibition regime.

As r$_{co} >$ r$_m$ the pulsar can be expected to be outside the
centrifugally inhibited regime and consequently should exhibit X-ray
pulsations. However, our estimates of the magnetospheric and
corotation radii are within a factor of two indicating that the source
may have recently transitioned from the centrifugally inhibited
regime. However, there are a number of uncertainties in the
calculation of the magnetospheric radius; a larger distance or an
$\eta <$1 would imply a higher mass accretion rate and reduce the
estimate of r$_{m}$ as would a lower magnetic field strength.

IBIS/ISGRI clearly measures the spin period of the pulsar in the A0535+262
system as 103.7~$\pm$~0.1 seconds.  If we add this to previous
measurements of the spin period of the pulsar we see the long term
pulse period history shown in Figure~\ref{fig:pp_hist}. It is clear
that since the BATSE observations of \citet{Finger96} the pulsar
underwent a period of consistent spin-down, slowing from $\sim$103.2
seconds to $\sim$103.7 seconds by the IBIS/ISGRI measurement in 2003.  Based
upon these measurements we estimate a spin-down rate of
$\dot{P}$~$\sim$~1.5~$\times$~10$^{-9}$~s~s$^{-1}$.  Spin-down is to
be expected as the system has been in a period of inactivity with no
reported X-ray outbursts from 1994-2005; the lack of X-ray activity
implies that no substantial accretion disk has formed around the
pulsar to provide any accretion torques to spin-up the pulsar.  The
Type II outburst in May/June 2005 followed by the type I outburst in
August/September 2005 spin the pulsar up to a period of
P~=~103.39315~$\pm$~0.00005~s \citep{Caballero2007}.

From the magnetic field strength we calculate the expected spin-down
through magnetic dipole losses as 9.4~$\times$~10$^{-16}$s~s$^{-1}$,
this is many orders of magntitude less than the observed spin-down and
hence cannot be the cause of the change in the pulsar period.
Additionally, the spin-down is of too high a magnitude to be explained
by the ejection of matter by the neutron star.  Consquently, the
spin-down of the pulsar must be the result of torques imparted from an
accretion disk around the neutron star; this implies that the
existence of a disk around the neutron star is not restricted to
periods immediately surrounding Type I or Type II outbursts.
 
\section{Conclusions}
The results presented here examine the high energy emission from
A0535+262 during a period in which the source was believed to be
quiescent.  The source is observed to be transient in the 18--100 keV
energy band by INTEGRAL.  The source luminosity is completely incompatible with Type II outbursting behaviour, is at the low end for a Type I outburst and is at the high end for quiescent emission.  The outburst occurs $\sim$21 days prior to periastron.  Although it may be a Type I outburst we examine the possibilty that the source activity is attributable to a flare during the quiescent phase.  \citet{Coe2006} report that during the flare the Be star circumstellar disk was not of sufficent size to support a Type I outburst.  The flaring nature of the source may be attributable to a brief period of increased mass transfer.  We suggest that this increased mass transfer could originate in a period of increased stellar wind density from the Be donor star although a low level Type I outburst cannot be completely ruled out.

Measurements of the spin period of the pulsar indicate that the source
has continued the spin-down trend which has been observed in the
source since it entered the quiescent phase in 1994.  A pulse period
of 103.7~$\pm$~0.1 seconds is longer than any other period estimates
made since 1994 and suggests a continuous process of spin-down at a
rate of $\dot{P}$~$\sim$1.5~$\times$~10$^{-9}$~s~s$^{-1}$.  Such a
level of spin-down can only be achieved through torquing drag processes
which implies the regular presence of a residual accretion disk around the
pulsar.

A0535+262 is a well observed system during periods of typical Type I and Type
II outbursts, however few measurements have been made of it during its
quiescent phases.  The INTEGRAL observations imply that during the
quiescent phase this system remains active and, by the nature of the
detection, is highly variable.  Extrapolating the IBIS/ISGRI spectrum using
the HEASARC \emph{webPIMMS} tool estimates that the count rate in the
RXTE-ASM would have been $\sim$2.4 counts s$^{-1}$ had it observed the
source.  However, examining the ASM observations of A0535+262 in the
week before and after the INTEGRAL observations indicate that the
light curve had a mean of 0.75 and a standard deviation of 2.4 counts
s$^{-1}$.  Consequently, it is unlikely that had the ASM observed
A0525+262 at the time of the flare the instrument sensitivity would
have been sufficient to identify the activity.

\section*{Acknowledgments}
Based on observations with INTEGRAL, an ESA project with instruments
and science data centre funded by ESA member states (especially the PI
countries: Denmark, France, Germany, Italy, Switzerland, Spain), Czech
Republic and Poland, and with the participation of Russia and the USA.

We acknowledge funding via PPARC grant PP/C000714/1.

\label{lastpage}

\end{document}